\documentclass[aps,prl,showpacs,preprintnumbers,twocolumn]{revtex4-1}
\usepackage{geometry}                
\usepackage{graphicx}
\usepackage{amssymb}
\usepackage{dsfont}
\usepackage{epstopdf}
\usepackage{color}
\usepackage{mathtools}
\usepackage{hyperref}

\definecolor{red}{rgb}{1,0,0}
\definecolor{blue}{rgb}{0,0,1}
\definecolor{black}{rgb}{0,0,0}
\newcommand{\blue}{}

\newcommand{\p}{\partial}
\newcommand{\eq}[1]{\begin{align}#1\end{align}}

\newcommand{\ffrac}[2]{\mbox{$\frac{#1}{#2}$}}
\newcommand{\half}{\mbox{$\frac{1}{2}$}}

\newcommand{\tr}{\mbox{tr}}

\usepackage{scalerel,stackengine}
\stackMath
\newcommand\widecheck[1]{%
\savestack{\tmpbox}{\stretchto{%
  \scaleto{%
    \scalerel*[\widthof{\ensuremath{#1}}]{\kern-.6pt\bigwedge\kern-.6pt}%
    {\rule[-\textheight/2]{1ex}{\textheight}}
  }{\textheight}%
}{0.5ex}}%
\stackon[1pt]{#1}{\scalebox{-1}{\tmpbox}}%
}

\newcommand{\alphab}{\hat{\alpha}}

\newcommand{\Dpsi}{\mathcal{D}\psi}

\newcommand{\Lc}{\mathcal{L}}
\newcommand{\Ac}{\mathcal{A}}
\newcommand{\rv}{\vec{r}}

\newcommand{\nv}{\vec{n}}

\newcommand{\uv}{\vec{u}}
\newcommand{\kv}{\vec{k}}

\newcommand{\rvp}{\vec{r}\;'}

\newcommand{\PP}{\mathbb{P}}

\newcommand{\Psib}{\hat\Psi}
\newcommand{\delb}{\hat\delta}
\newcommand{\sigmab}{\hat\sigma}
\newcommand{\epsb}{\hat\epsilon}

\newcommand{\sigmabar}{{\hat{\overline \sigma}}}

\newcommand{\fsl}[1]{\ensuremath{\mathrlap{\!\not{\phantom{#1}}}#1}}

\begin{document}
\title{Field theory for amorphous solids}
\author{E. DeGiuli}
\affiliation{Institut de Physique Th\'eorique Philippe Meyer, \'Ecole Normale Sup\'erieure, \\ PSL University, Sorbonne Universit\'es, CNRS, 75005 Paris, France}

\begin{abstract}
Glasses at low temperature fluctuate around their inherent states; glassy anomalies reflect the structure of these states. Recently there have been numerous observations of long-range stress correlations in glassy materials, from supercooled liquids to colloids and granular materials, but without a common explanation. Herein it is shown, using a field theory of inherent states, that long-range stress correlations follow from mechanical equilibrium alone, with explicit predictions for stress correlations in 2 and 3 dimensions. `Equations of state' relating fluctuations to imposed stresses are derived, as well as field equations that fix the spatial structure of stresses in arbitrary geometries. Finally, a new holographic quantity in 3D amorphous systems is identified. 
\end{abstract}
\maketitle

The low temperature properties of solids necessarily reflect their inherent states, and the local neighbourhoods thereof. In Debye's model, applicable to crystals, inherent states are perfect crystalline arrangements, and harmonic vibrations are phonons. In constrast, for amorphous solids, there is no accepted, simple description of inherent states. Since glasses universally present thermal and vibrational anomalies with respect to crystals \cite{Phillips81,Pohl02}, for example in their heat capacity and thermal conductivity, one might hope that a simple description would be forthcoming. Moreover, even out-of-equilibrium amorphous solids such as granular materials, emulsions, and colloids present similar phenomenology in their vibrational properties \cite{Seguin16,Lin16b,Zargar17,Zhang17}, further suggesting a unified approach. 

Recent observations of stress correlations support such unification. In simulations both of granular materials \cite{Henkes09a,Wu17} and deeply supercooled liquids  \cite{Lemaitre14,Wu15,Lemaitre15}, the spatial shear-stress correlator has quadrapolar anisotropy and a power-law decay $\propto 1/r^d$ in $d$ dimensions. Similar observations have been made for strain correlations in experiment, both for colloids \cite{Jensen14,Illing16} and granular materials \cite{Le-Bouil14}. It was argued in \cite{Lemaitre14,Lemaitre15,Maier17} that such correlations could be explained by the dynamical process by which the systems evolve, namely by elastic relaxation of so-called Eshelby transformations \cite{Lemaitre14,Lemaitre15}, or by a mode-coupling approach \cite{Maier17}. However, in granular materials the elastic range is extremely small, such that essentially all observed deformation is plastic \cite{Roux10,Schreck11}, casting doubt on these dynamical explanations. Very recently, Lema\^itre has shown that in 2D systems, mechanical equilibrium (ME) and material isotropy are sufficient to explain the anisotropy of stress correlations \cite{Lemaitre17}, consistent with the theory of Henkes and Chakraborty for granular materials \cite{Henkes09a}, also in 2D. 
Since 2D solids must be unusual, by the Mermin-Wagner theorem, it is imperative to see if these results survive in 3D. A general theory of inherent states should predict these stress correlations in both 2D and 3D and, ideally, be applicable both to glasses and athermal systems. 

Inherent states are defined by conditions of ME. Unlike crystals, for which these constraints are trivially satified by symmetry, in amorphous materials ME requires a delicate balance among the microscopic degrees of freedom.
 This is most spectacularly displayed near the jamming transition at which rigidity is lost altogether \cite{Liu10,Wyart12}, but it remains an organizing principle throughout the solid phase. In this work we present a statistical field theory of inherent states, using only general principles valid at large probing scales (Fig 1). We show that ME alone predicts the full form of stress correlations in 2D and 3D, and derive the equations of state relating fluctuations to imposed stresses. We also find field equations that determine the stress field in arbitrary domains. 

In this Letter, we consider solids with both attractive and repulsive interactions, and present the main physical results of potential interest to the glass community. In an accompanying manuscript \cite{atmp_DeGiuli18_long}, we show calculational details and extend the results to solids in which forces are strictly repulsive. 
 In that work we discuss our theory in the context of previous work on so-called Edwards ensembles \cite{Bi15}, discussed in the granular matter community. 

Amorphous systems are controlled either by external stress, or by the particle density; in this work we consider the stress ensemble \cite{Henkes09a,Chakraborty10}, assuming that geometric degrees-of-freedom (DOF) have been marginalized over. 

\begin{figure}[t!]
\includegraphics[width=1\columnwidth,viewport = 30 250 610 510,clip]{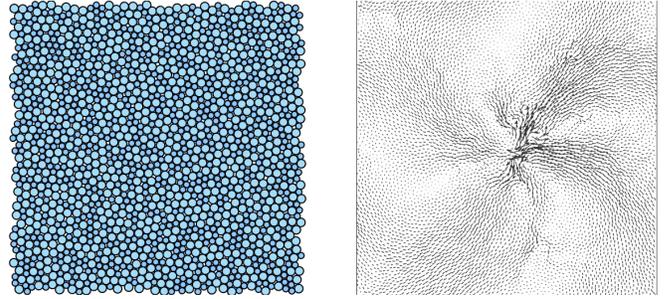}
\caption{ Illustration of response to a localized force dipole in a model glass, courtesy of E. Lerner \cite{Lerner18}. Plotted on the right is $|\rv| |\vec{u}|$, where $\vec{u}$ is the displacement response, decaying approximately as $\sim 1/|\rv|$, where $|\rv|$ is distance to the source. Beyond a few particle diameters, the response is well-represented by a disordered, continuum field \cite{Goldenberg07}. In field theory, the glass (left) is replaced by a continuum, whose structure is characterized by one or several smooth structural fields. 
}\label{fig1}
\end{figure} 

{\bf Stress Correlators:} We first show how 
 ME constrains the tensorial form of correlation functions, without additional hypotheses. This has recently been vividly demonstrated by Lema\^itre in 2D systems \cite{Lemaitre17}; here we show how a gauge formulation of the problem immediately gives a compact and complete answer, and then generalize to 3D systems. 

In the absence of body forces, the stress tensor of a system in ME must be symmetric, $\sigmab=\sigmab^t$, from torque balance, and solenoidal, $0=\nabla \cdot \sigmab$, from force balance. These relations imply that the stress tensor is both left-transverse, $k_i \sigma_{ij}=0$ for any wavevector $k_i$, and right-transverse, $\sigma_{ij} k_j=0$. The number of DOF in $\sigmab$ is reduced from $d^2$ in $d$ dimensions down to $d(d-1)/2$, and thus metastable states, as characterized by their stress, occupy a vanishing fraction of configuration space. 

Instead of carrying these constraints along in all theoretical manipulations, it would be extremely convenient to work directly on the manifold of metastable states. This is accomplished by a gauge representation in which the constraints are identically satisfied.  In two dimensions, this is achieved by \eq{ \label{airy}
\sigmab = \nabla \times \nabla \times \psi, \qquad \sigma_{ik} = \epsilon_{ij} \epsilon_{kl} \p_j \p_l \psi,
}
where $\psi$ is known as the Airy stress function \cite{Muskhelishvili63} and $\epsilon_{12}=-\epsilon_{21}=1, \epsilon_{11}=\epsilon_{22}=0$. It is easily verified that for any function $\psi(\rv)$, both $\sigmab=\sigmab^t$ and $0=\nabla \cdot \sigmab$ are identically satisfied. Moreover, this representation also exists at the particle scale \cite{Ball02,Henkes09,DeGiuli11,DeGiuli14a}. The price of the gauge representation is that stresses are invariant under the gauge transformation $\psi \to \psi + \vec{a} \cdot \rv + b$, for any constants $\vec{a}$ and $b$. Stresses depend only on the curvature of $\psi$. 

In using $\psi$, we put ourselves on the manifold of metastable states, in 2D. It follows that in any ensemble the fundamental correlation function is then
\eq{
C_\psi(\rv,\rvp) = \langle \psi(\rv) \psi(\rvp) \rangle - \langle \psi(\rv) \rangle \langle \psi(\rvp) \rangle,
}
which is invariant under gauge transformations. The stress-stress correlation function is
$\langle \sigma_{ij}(\rv) \sigma_{kl}(\rvp) \rangle_c = \epsilon_{im} \epsilon_{jn} \epsilon_{kp} \epsilon_{lq} \p_m \p_n \p'_p \p'_q C_\psi(\rv,\rvp)$
and, assuming homogeneity, can be written
\eq{ \label{corr3}
\langle \sigma_{ij}(\rv) \sigma_{kl}(0) \rangle_c & = \epsilon_{im} \epsilon_{jn} \epsilon_{kp} \epsilon_{lq} \p_m \p_n \p_p \p_q C_\psi(\rv,0).
}
Since the pressure-pressure correlator is $\langle p(\rv) p(0) \rangle_c = \ffrac{1}{4} \nabla^4 C_\psi(\rv,0)$, in periodic systems the full correlation function at finite-wavevector can be written
\eq{ \label{corr4}
\langle \sigma_{ij}(\kv) \sigma_{kl}(-\kv) \rangle_c = 4 P_{ij}^T P_{kl}^T \langle p(\kv) p(-\kv) \rangle_c,
}
where $P_{ij}^T = \delta_{ij} - k_i k_j |k|^{-2}$ is the transverse projector \cite{DiDonna05}. 
Eq.\ref{corr4} holds even in anisotropic systems. 

From \eqref{corr3} one can easily determine all components of the stress correlator if $C_\psi(\rv,0)$ is known, which will be derived below. 
First, we extend this result to 3D; the analogous gauge representation is 
\eq{ \label{beltrami}
\sigmab = \nabla \times \nabla \times \Psib, \qquad \sigma_{il} = \epsilon_{ijk} \epsilon_{lmn} \p_j \p_m \Psi_{kn},
}
where $\Psib$, a symmetric second-order tensor, is the Beltrami stress tensor \cite{Gurtin63}. Note that, by convention, the tensor curl is defined by acting on the right-most index, i.e. $(\nabla \times \Psib)_{ij} = \epsilon_{ikl} \p_k \Psi_{jl}$. One easily verifies that for any tensor field $\Psib(\rv)$, \eqref{beltrami} describes a symmetric, solenoidal stress tensor. A discrete representation of $\Psib$ also exists \cite{DeGiuli11}. Since $\Psib$ has the same number of DOF as $\sigmab$, we infer that some of these must be redundant. Indeed, for any vector field $\vec{p}(\rv)$, the stress tensor is invariant under the transformation $\Psib \to \Psib + \nabla \vec{p} + (\nabla \vec{p})^t$,
a nontrivial gauge freedom \cite{Wang13a}. Accordingly, $\Psib$ 
can be further reduced. We will use the Maxwell gauge $\Psi_{ij} = \delta_{ij} \psi_j$ (no sum on $j$), which reduces the number of DOF in $\Psib$ from $6$ to $3$, as required. 
 Then the fundamental correlation function is
\eq{
C_{ij}(\rv,\rvp) = \left\langle \psi_i(\rv) \psi_j(\rvp) \right\rangle_c,
}
which has at most 6 independent components, and is gauge invariant. If isotropy and homogeneity are assumed, then this has two independent components, $A(\rv)=C_{ii}(\rv,0)$ (no sum on $i$) and $B(\rv)=C_{ij}(\rv,0)$ ($i\neq j$). One easily sees that all stress correlators involving longitudinal components vanish\cite{atmp_DeGiuli18_long}, so that again only the transverse-transverse stress correlator survives, which now, however, is tensorial. 

Although simple to derive in the gauge formulation, the above results completely prescribe the tensorial structure of the stress correlator, a major aim of previous works \cite{Lemaitre14,Lemaitre15,Maier17,Lemaitre17} . We also see that material isotropy is not important in determining this structure, although it would simplify the implied derivatives. To obtain predictions for the correlation functions, we now proceed to field theory.

{\bf Gauge field theory of inherent states: } Since our interest is in properties at large probing length, we work in the continuum (Fig. \ref{fig1}). We are interested both in glasses and out-of-equilibrium athermal systems. For glasses, the probability distribution over inherent states will contain a Gibbs contribution from the energy at the glass transition temperature, but also a nontrivial entropic contribution, the `complexity' \cite{Berthier11b}. For athermal systems, we do not even have a Gibbs contribution from which to begin a theory.  


To construct a stress ensemble valid out of equilibrium, we take an operational point of view: typically, one can probe a system only through forcing at the boundary. Unlike thermally equilibrated systems, an athermal ensemble needs to be explicitly explored through systematic forcing. Such an ensemble can be explored dynamically, as in quasistatic shear flow in a Couette cell, but we need not restrict ourselves to this setting; indeed, most numerical simulations and experiments generate an ensemble simply by repeated application of a preparation protocol. 
 In order for a variable to be controllable under an athermal ensemble generated by boundary forcing, it must be {\it holographic,} that is, determined by boundary quantities only. The stress tensor in a mechanically equilibrated system is indeed such a quantity, as is easily seen \cite{atmp_DeGiuli18_long}. 
 

In addition to being holographic, controllable quantities should be additive, so that the thermodynamic limit can exist. Thus the true controllable quantity is $\int_\Omega \sigmab$, known as the force-moment tensor. It is a surprising fact that in addition to $\int_\Omega \sigmab$, there is another holographic, additive quantity depending on the stress. To see this, we initially consider 2D, and use 
\eqref{airy}. 
 One sees that the determinant of the stress tensor is 
\eq{
\det \sigma & = \half \epsilon_{ij} \epsilon_{kl} \sigma_{ik} \sigma_{jl} 
 = \half \p_j \big( (\p_l \psi) \sigma_{jl} \big)
}
where we used $\epsb^T \cdot \epsb = \delb$ and $\nabla \cdot \sigmab = 0$. 
{\blue Thus, using the divergence theorem,} $\Ac = \int_\Omega \det \sigma$ can be written as a boundary quantity. It can be shown that if $\nv\cdot\sigmab$ is known around the boundary, where $\nv$ is a boundary normal, then $\Ac$ is also fixed.
In previous work, the discrete quantity corresponding to $\Ac$ has been called the Maxwell-Cremona area \cite{Tighe08,Tighe10a,Tighe10b,DeGiuli11,Wu15a}. 

Let us now show that a similar quantity also exists in 3D, although to our knowledge it has never been reported before. Using \eqref{beltrami}, simple algebra shows that the determinant of $\sigmab$ is 
\eq{
\det \sigma & = \frac{1}{3}  \p_p \left[ \epsilon_{lmn} (\nabla \times \Psib)_{lq} \sigma_{pm} \sigma_{qn} \right],
}
a total divergence. The quantity $\Ac = \int_\Omega \det \sigma$ could be called the Beltrami volume. 
It can be shown, similar to the 2D case, that $\Ac$ can be recovered if $\hat{n} \cdot \sigmab$ is known around a closed boundary.

Having identified controllable quantities $\int_\Omega \sigma$ and $\Ac$, we can construct a canonical ensemble in which the control parameters are temperature-like variables conjugate to $\sigma$ and $\Ac$. This leads to an action
\eq{
S_0 = \int_{\Omega} dV \; \left[ \alphab : \sigmab + \gamma \det \sigmab \right],
}
where $\alphab^{-1}$ has been called the angoricity \cite{Blumenfeld09}, and $\gamma$ has been called the keramicity \cite{Bililign18}. The justification for the canonical ensemble is based upon an assumed factorization of the probability distribution for macroscopic variables into that of subsystems, as discussed in detail in Refs. \cite{Bertin06,Henkes07,Henkes09,Wu15a}, and has been successfully tested in experiments on granular matter \cite{Puckett13,Bililign18}. In such a generalized Gibbs ensemble, the temperature-like variables $\alphab$ and $\gamma$ are argued to be spatially constant \cite{Bertin06}.

To complete the specification of the probability distribution of the stress field, we need to address (i) the hard constraints necessary to impose ME, and (ii) the {\it a priori} probability with which each metastable state is sampled. 
As discussed above, we can efficiently work on the manifold of metastable states by writing $\sigmab$ as a functional of $\psi$ (2D) and $\psi_i$ (3D). This leads to
\eq{ \label{p1}
\PP[\sigmab[\psi]] = \frac{1}{Z} \omega[\sigmab[\psi]] e^{-S_0[\psi]},
}
where $\omega$ is the sampling probability of the state defined by $\sigmab[\psi]$, and we use $\psi$ to refer either the scalar Airy stress function (2D) or its vectorial analog in the Maxwell gauge (3D). It is implicit that in $\psi-$space there is a large-wavenumber ultraviolet cutoff $\Lambda \propto 1/D$, where $D$ is the typical particle diameter.

In a strict canonical ensemble, the sampling probability $\omega$ would be unity, as was taken in previous work on the stress ensemble \cite{Henkes07,Henkes09}. 
 In fact, there is no general justification for the flat measure out of equilibrium, even if it was observed to hold to a good approximation in several model systems \cite{Barrat00,Biroli01a}. In general, we expect the flat measure to be unrealistic for a simple reason: since $S_0$ can be written in terms of boundary quantities only, if $\omega\equiv 1$ then \eqref{p1} would be invariant under arbitrary diffeomorphisms in the bulk, limited only by the UV cutoff $\Lambda$. This would allow arbitrarily wild fluctuations of the field down to the scale $\Lambda$, which is not physical: a solid stores elastic energy, and whenever elasticity is present, stress fluctuations will be penalized. 

The sampling probability $\omega$ must thus be nontrivial. For arbitrary 
 $\omega$ nothing can be computed, but we are rescued by the continuum limit. The general theory of the renormalization group indicates that when a system is probed at long length scales, most of its microscopic details are irrelevant \cite{Zee10,Zinn-Justin96}. {\blue For any Lagrangian theory, power counting can be applied to see which terms are necessary to retain in a Landau-Wilson expansion}
\eq{ \label{om1}
\omega[\sigma[\psi]] = e^{-\int dV [ A_1[\sigmab] + A_2[\sigmab,\sigmab] + \ldots ]},
}
where each $A_i$ is a differential operator linear in each argument.

We need to consider the symmetry properties of the stress tensor. In systems with both repulsive and attractive interactions, a term linear in stress, which is not invariant under $\sigmab \to -\sigmab$, will not ensure a well-behaved distribution; for this a term quadratic in stress is necessary. In the continuum limit, the lowest order term necessary to tame fluctuations is then $\eta \;\sigmab : \sigmab$. We assume that $\eta$ defines the correct units in which to construct the field theory, so that if lengths have dimension $+1$ then $\sigmab$ has canonical dimension $-d/2$ to make the action dimensionless. A term of the form $\p^n \sigma^q$ then has a coupling constant with operator dimension $\delta_{n,q}=d-n-qd/2$. Relevant operators are those with $\delta_{n,q} \geq 0$ \cite{Zee10}. In $d=2,3$ this includes only $q=1,n \leq1$, and $q=2,n=0$. Assuming reflection symmetry, so that a term $g_{ijk} \p_i \sigma_{jk}$ is excluded, the only new isotropic terms added are $\eta \;\tr^2\sigmab$ and $g \;\tr\; \sigmab^2$. Under strongly anisotropic forcing further terms would be necessary \cite{atmp_DeGiuli18_long}.

We are thus led to consider
\eq{ \label{p2}
\PP[\sigmab[\psi]] = \frac{1}{Z} e^{-S[\psi]}, \;\; S = \int_{\Omega} dV \; \Lc[\psi], 
}
with
\eq{ \label{l1}
\Lc[\psi] & = \alphab : \sigmab + \gamma \det \sigmab + \half \eta \; \tr^2 \sigmab + \half g \; \tr\;\sigmab\cdot \sigmab.
}
As usual, it is sufficient to compute $Z = \int \Dpsi \;e^{-S}$ to extract the behavior of controllable quantities. 

At this stage, it is clear that we could have arrived at \eqref{l1} with power counting alone, without any consideration of controllable quantities. However, this would miss an important point: the parameters $\alphab$ and $\gamma$ are conjugate to holographic quantities, and hence under experimental control. The `elastic' parameters $\eta$ and $g$ instead reflect the properties of the particles and should not depend on details of the experimental protocol. We will see the importance of this distinction below. 

{\bf Results. 2D: } The computation of $Z$ is detailed in \cite{atmp_DeGiuli18_long}. Here we emphasize 3 key results. First, we obtain the equation of state 
\eq{
\sigmabar = \frac{1}{\gamma-g} \alphab - \frac{(\eta+\gamma)}{(\gamma-g)(\gamma + g + 2 \eta)} \delb \;\tr \;\alphab,
}
relating the temperature-like quantities $\alphab$ and $\gamma$ to the mean stress $\langle \sigmab \rangle = \sigmabar$. Second, we obtain the field equation $\nabla^4 \psi = 0$ that governs the distribution of stress in an arbitrary geometry, to be solved along with boundary conditions given in \cite{atmp_DeGiuli18_long}. Finally, we obtain the stress correlator by solving $\nabla^4 \psi_g(\rv) = -\tilde\eta^{-1} \nabla \nabla : \alphab_g$, with $\tilde\eta=\eta+g$ and $\alphab_g = \alphab_0 \delta(\rv-\rvp)$, and using
\eq{ \label{corr5}
\langle \sigmab(\rv) \sigmab(\rvp) \rangle_c = -\half \delta \sigmab_g(\rv) / \delta \alphab_g(\rvp) -\half \delta \sigmab_g(\rvp) / \delta \alphab_g(\rv), 
}
evaluated at $\alphab_g=0$. In an infinite domain, the solution to a source at the origin is
\eq{
4 \pi \tilde\eta \; \psi_g = -\alpha \log r^2 + a \cos 2\theta + b \sin 2\theta, 
}
where 
$\alphab_0 = \begin{pmatrix} \alpha + a & b \\ b & \alpha-a \end{pmatrix}$.
 Boundary conditions can be applied by adding to $\psi_g$ an appropriate biharmonic function $\psi_b$, $\nabla^4 \psi_b=0$. Explicitly, the pressure-pressure correlator is
\eq{
\langle p(\rv) p(0) \rangle_c = (4\tilde\eta)^{-1} \; \delta(\rv)
}
This is short-range, but all second derivatives will have a $1/r^2$ decay with appropriate anisotropic dependencies, following Eq.\eqref{corr4}. Note that the prediction of a perfect $\delta(\rv)$ correlator is an artifact of the truncation of $\Lc$ to Gaussian order; if higher order terms were included in $\Lc$, such as $\tr^4 \sigma$, then the pressure-pressure correlator would have an exponential decay over the particle size  length scale $\sim D$, as found in \cite{Henkes09a}. The above results correspond to $C_\psi(\rv,0) = r^2 \log r /(4\pi \tilde\eta)$, from which all correlators can be obtained.

{\bf Results. 3D}: From the computation of $Z$ \cite{atmp_DeGiuli18_long}, we find that the mean stress $\sigmabar = \nabla \times \nabla \times \overline{\Psi}$ is fixed by the equation of state
\eq{
0 = \alpha_{ij} + \eta \; \delta_{ij} \sigmabar_{kk} + g \; \sigmabar_{ij} + \half \gamma \; \epsilon_{ikl} \epsilon_{jmn} \sigmabar_{km} \sigmabar_{ln},
}
which is now nonlinear, owing to the $\det \sigma$ term. Second, we find that stresses in arbitrary geometries can be found by solving 
\eq{ \label{beltramimichell}
0 = (\eta+g) \p_i \p_j \sigma_{kk} + g \nabla^2 \sigma_{ij},
}
subject to appropriate boundary conditions. This is equivalent to the Beltrami-Michell equation of linear elasticity \cite{Sadd09}, with an effective Poisson ratio $\nu = -\eta/(\eta+g)$. Finally, the stress correlator is again determined by \eqref{corr5}, where in an infinite domain the solution to a source at the origin is now  
\eq{
g \sigmab_g = -\alphab_g - \frac{2\eta}{\tilde g} \delb \; \nabla \cdot \uv + \frac{\eta}{\tilde g} \tr \;\alpha_g + \nabla \uv + (\nabla \uv)^t,
}
with
\eq{
\nabla \cdot \uv & = -\frac{1}{8\pi} \frac{\tilde g}{2\eta+g} \left[ \alphab_0 : \nabla \nabla - \frac{\eta \;\tr\;\alphab_0}{\tilde g} \nabla^2 \right] \frac{1}{r}, \\
\uv & = - \frac{1}{4\pi} \left[ \alphab_0 - \frac{\eta}{\tilde g} \delb \; \tr \;\alphab_0 \right] \cdot \nabla \frac{1}{r}  \notag \\
& \qquad - \frac{\eta+g}{\tilde g} \frac{1}{\nabla^2} \nabla \nabla \cdot \uv.
}
and $\tilde g = 3\eta+g$. The complex tensorial structure resulting from this solution precisely matches what was found in Ref. \cite{Lemaitre15}. For example, the isotropic part is
\eq{
\tr\; \sigmab_g = -\frac{2 \alpha}{2\eta+g} \delta(\rv) - \frac{1}{2\eta+g} \fsl{\alphab} : \nabla \nabla \frac{1}{4\pi r},
}
where $\alphab_0 = \alpha \delb + \fsl{\alphab}$ with $\tr \; \fsl{\alphab} = 0$. The pressure-pressure correlator is short-range, while the pressure-shear correlator has anisotropy and long-range decay determined by the Oseen tensor $\nabla \nabla r^{-1}$.

{\bf Holography: } From the above results we can see that the holographic terms play a fundamentally different role from the others. 
 Indeed, in large systems $\alphab$ and $\gamma$ appear only in the equations of state, so that they 
 control the system-spanning $\kv=0$ fluctuations, but not the finite wavevector $|\kv|>0$ fluctuations. As a result the stress-stress correlation function should have a discontinuity or kink at $\kv=0$.

To see these distinct fluctuations, let $\overline{x} \equiv \ffrac{1}{|\Omega|} \int_\Omega dV \; x(\rv)$ denote a spatial average, and consider
\eq{
C_e = \left\langle \left(\overline{p - \langle p \rangle} \right)^2 \right\rangle, \qquad C_s = \left\langle \overline{ (p - \overline{p})^2 } \right\rangle,
} 
where $\langle \; \rangle$ denotes an ensemble average. $C_e$ measures the ensemble pressure fluctuations while $C_s$ measures spatial pressure fluctuations. In 2D we find (Appendix 4)
\eq{
C_e = \frac{1}{2 V (2\eta+g+\gamma)}, \qquad C_s = \frac{\Lambda^2}{16 \pi \tilde\eta} - C_e
}
Curiously, the total fluctuations $C_e+C_s$ are fixed by $\tilde\eta$ only, while the ensemble fluctuations depend additionally on $\gamma$. This explains the discontinuity at $\kv=0$ observed in Ref \cite{Lemaitre17}.



{\bf Conclusion: } {\blue Using general field-theoretical arguments, we have derived a field theory for the stress ensemble of athermal amorphous solids, and derived explicit forms of the long-range stress correlations in both 2D and 3D. }
 The main assumptions underlying the theory are that: (i) all quantities are probed at lengths much larger than the particle size; (ii) all interactions between the stresses are themselves local; and (iii) no strict constraints such as positivity or Coulomb friction have been imposed on the forces; this last assumption is relaxed in \cite{atmp_DeGiuli18_long}, where important modifications to the present theory for strictly repulsive interactions are shown. Furthermore, we derived equations relating fluctuations to imposed stresses, and field equations that fix the spatial form of stresses in arbitrary domains. We also identified a new holographic quantity in 3D systems. 

{\blue Our analysis has been restricted to the athermal limit. At finite temperature, stress correlators will receive an additional thermal contribution, proportional to $T$, and only involving longitudinal stress components. This is not expected to affect the transverse-transverse correlations discussed here; a schematic construction of such a term is shown in Supplementary Information. More importantly, at $T>0$ glasses can transition between IS's through activated processes. In \cite{atmp_DeGiuli18_long}, we show that the theory as written here has an infinite-dimensional symmetry; we expect transitions between IS's to occur along the action of this symmetry. The effect of activated processes will then depend how this symmetry is lifted; we expect this to be important also for plasticity and yielding, to be tackled in future work.

}




\begin{acknowledgments}
The author kindly acknowledges many exchanges within the Simons `Cracking the glass' collaboration, especially frequent discussions with E. Lerner.
\end{acknowledgments}

\bibliographystyle{apsrev4-1}
\bibliography{../Glasses} 

\begin{thebibliography}{48}%
\makeatletter
\providecommand \@ifxundefined [1]{%
 \@ifx{#1\undefined}
}%
\providecommand \@ifnum [1]{%
 \ifnum #1\expandafter \@firstoftwo
 \else \expandafter \@secondoftwo
 \fi
}%
\providecommand \@ifx [1]{%
 \ifx #1\expandafter \@firstoftwo
 \else \expandafter \@secondoftwo
 \fi
}%
\providecommand \natexlab [1]{#1}%
\providecommand \enquote  [1]{``#1''}%
\providecommand \bibnamefont  [1]{#1}%
\providecommand \bibfnamefont [1]{#1}%
\providecommand \citenamefont [1]{#1}%
\providecommand \href@noop [0]{\@secondoftwo}%
\providecommand \href [0]{\begingroup \@sanitize@url \@href}%
\providecommand \@href[1]{\@@startlink{#1}\@@href}%
\providecommand \@@href[1]{\endgroup#1\@@endlink}%
\providecommand \@sanitize@url [0]{\catcode `\\12\catcode `\$12\catcode
  `\&12\catcode `\#12\catcode `\^12\catcode `\_12\catcode `\%12\relax}%
\providecommand \@@startlink[1]{}%
\providecommand \@@endlink[0]{}%
\providecommand \url  [0]{\begingroup\@sanitize@url \@url }%
\providecommand \@url [1]{\endgroup\@href {#1}{\urlprefix }}%
\providecommand \urlprefix  [0]{URL }%
\providecommand \Eprint [0]{\href }%
\providecommand \doibase [0]{http://dx.doi.org/}%
\providecommand \selectlanguage [0]{\@gobble}%
\providecommand \bibinfo  [0]{\@secondoftwo}%
\providecommand \bibfield  [0]{\@secondoftwo}%
\providecommand \translation [1]{[#1]}%
\providecommand \BibitemOpen [0]{}%
\providecommand \bibitemStop [0]{}%
\providecommand \bibitemNoStop [0]{.\EOS\space}%
\providecommand \EOS [0]{\spacefactor3000\relax}%
\providecommand \BibitemShut  [1]{\csname bibitem#1\endcsname}%
\let\auto@bib@innerbib\@empty
\bibitem [{\citenamefont {Anderson}(1981)}]{Phillips81}%
  \BibitemOpen
  \bibfield  {author} {\bibinfo {author} {\bibfnamefont {A.}~\bibnamefont
  {Anderson}},\ }\href@noop {} {\emph {\bibinfo {title} {Amorphous Solids: Low
  Temperature Properties}}},\ edited by\ \bibinfo {editor} {\bibfnamefont
  {W.~A.}\ \bibnamefont {Phillips}},\ \bibinfo {series} {Topics in Current
  Physics}, Vol.~\bibinfo {volume} {24}\ (\bibinfo  {publisher} {Springer,
  Berlin},\ \bibinfo {year} {1981})\BibitemShut {NoStop}%
\bibitem [{\citenamefont {Pohl}\ \emph {et~al.}(2002)\citenamefont {Pohl},
  \citenamefont {Liu},\ and\ \citenamefont {Thompson}}]{Pohl02}%
  \BibitemOpen
  \bibfield  {author} {\bibinfo {author} {\bibfnamefont {R.~O.}\ \bibnamefont
  {Pohl}}, \bibinfo {author} {\bibfnamefont {X.}~\bibnamefont {Liu}}, \ and\
  \bibinfo {author} {\bibfnamefont {E.}~\bibnamefont {Thompson}},\ }\href@noop
  {} {\bibfield  {journal} {\bibinfo  {journal} {Reviews of Modern Physics}\
  }\textbf {\bibinfo {volume} {74}},\ \bibinfo {pages} {991} (\bibinfo {year}
  {2002})}\BibitemShut {NoStop}%
\bibitem [{\citenamefont {Seguin}\ and\ \citenamefont
  {Dauchot}(2016)}]{Seguin16}%
  \BibitemOpen
  \bibfield  {author} {\bibinfo {author} {\bibfnamefont {A.}~\bibnamefont
  {Seguin}}\ and\ \bibinfo {author} {\bibfnamefont {O.}~\bibnamefont
  {Dauchot}},\ }\href@noop {} {\bibfield  {journal} {\bibinfo  {journal}
  {Physical review letters}\ }\textbf {\bibinfo {volume} {117}},\ \bibinfo
  {pages} {228001} (\bibinfo {year} {2016})}\BibitemShut {NoStop}%
\bibitem [{\citenamefont {Lin}\ \emph {et~al.}(2016)\citenamefont {Lin},
  \citenamefont {Jorjadze}, \citenamefont {Pontani}, \citenamefont {Wyart},\
  and\ \citenamefont {Brujic}}]{Lin16b}%
  \BibitemOpen
  \bibfield  {author} {\bibinfo {author} {\bibfnamefont {J.}~\bibnamefont
  {Lin}}, \bibinfo {author} {\bibfnamefont {I.}~\bibnamefont {Jorjadze}},
  \bibinfo {author} {\bibfnamefont {L.-L.}\ \bibnamefont {Pontani}}, \bibinfo
  {author} {\bibfnamefont {M.}~\bibnamefont {Wyart}}, \ and\ \bibinfo {author}
  {\bibfnamefont {J.}~\bibnamefont {Brujic}},\ }\href@noop {} {\bibfield
  {journal} {\bibinfo  {journal} {Physical Review Letters}\ }\textbf {\bibinfo
  {volume} {117}},\ \bibinfo {pages} {208001} (\bibinfo {year}
  {2016})}\BibitemShut {NoStop}%
\bibitem [{\citenamefont {Zargar}\ \emph {et~al.}(2016)\citenamefont {Zargar},
  \citenamefont {DeGiuli},\ and\ \citenamefont {Bonn}}]{Zargar17}%
  \BibitemOpen
  \bibfield  {author} {\bibinfo {author} {\bibfnamefont {R.}~\bibnamefont
  {Zargar}}, \bibinfo {author} {\bibfnamefont {E.}~\bibnamefont {DeGiuli}}, \
  and\ \bibinfo {author} {\bibfnamefont {D.}~\bibnamefont {Bonn}},\ }\href@noop
  {} {\bibfield  {journal} {\bibinfo  {journal} {EPL (Europhysics Letters)}\
  }\textbf {\bibinfo {volume} {116}},\ \bibinfo {pages} {68004} (\bibinfo
  {year} {2016})}\BibitemShut {NoStop}%
\bibitem [{\citenamefont {Zhang}\ \emph {et~al.}(2017)\citenamefont {Zhang},
  \citenamefont {Zheng}, \citenamefont {Wang}, \citenamefont {Zhang},
  \citenamefont {Jin}, \citenamefont {Hong}, \citenamefont {Wang},\ and\
  \citenamefont {Zhang}}]{Zhang17}%
  \BibitemOpen
  \bibfield  {author} {\bibinfo {author} {\bibfnamefont {L.}~\bibnamefont
  {Zhang}}, \bibinfo {author} {\bibfnamefont {J.}~\bibnamefont {Zheng}},
  \bibinfo {author} {\bibfnamefont {Y.}~\bibnamefont {Wang}}, \bibinfo {author}
  {\bibfnamefont {L.}~\bibnamefont {Zhang}}, \bibinfo {author} {\bibfnamefont
  {Z.}~\bibnamefont {Jin}}, \bibinfo {author} {\bibfnamefont {L.}~\bibnamefont
  {Hong}}, \bibinfo {author} {\bibfnamefont {Y.}~\bibnamefont {Wang}}, \ and\
  \bibinfo {author} {\bibfnamefont {J.}~\bibnamefont {Zhang}},\ }\href@noop {}
  {\bibfield  {journal} {\bibinfo  {journal} {Nature communications}\ }\textbf
  {\bibinfo {volume} {8}},\ \bibinfo {pages} {67} (\bibinfo {year}
  {2017})}\BibitemShut {NoStop}%
\bibitem [{\citenamefont {Henkes}\ and\ \citenamefont
  {Chakraborty}(2009)}]{Henkes09a}%
  \BibitemOpen
  \bibfield  {author} {\bibinfo {author} {\bibfnamefont {S.}~\bibnamefont
  {Henkes}}\ and\ \bibinfo {author} {\bibfnamefont {B.}~\bibnamefont
  {Chakraborty}},\ }\href {\doibase 10.1103/PhysRevE.79.061301} {\bibfield
  {journal} {\bibinfo  {journal} {Phys. Rev. E}\ }\textbf {\bibinfo {volume}
  {79}},\ \bibinfo {pages} {061301} (\bibinfo {year} {2009})}\BibitemShut
  {NoStop}%
\bibitem [{\citenamefont {Wu}\ \emph {et~al.}(2017)\citenamefont {Wu},
  \citenamefont {Karimi}, \citenamefont {Maloney},\ and\ \citenamefont
  {Teitel}}]{Wu17}%
  \BibitemOpen
  \bibfield  {author} {\bibinfo {author} {\bibfnamefont {Y.}~\bibnamefont
  {Wu}}, \bibinfo {author} {\bibfnamefont {K.}~\bibnamefont {Karimi}}, \bibinfo
  {author} {\bibfnamefont {C.~E.}\ \bibnamefont {Maloney}}, \ and\ \bibinfo
  {author} {\bibfnamefont {S.}~\bibnamefont {Teitel}},\ }\href@noop {}
  {\bibfield  {journal} {\bibinfo  {journal} {Physical Review E}\ }\textbf
  {\bibinfo {volume} {96}},\ \bibinfo {pages} {032902} (\bibinfo {year}
  {2017})}\BibitemShut {NoStop}%
\bibitem [{\citenamefont {Lema{\^\i}tre}(2014)}]{Lemaitre14}%
  \BibitemOpen
  \bibfield  {author} {\bibinfo {author} {\bibfnamefont {A.}~\bibnamefont
  {Lema{\^\i}tre}},\ }\href@noop {} {\bibfield  {journal} {\bibinfo  {journal}
  {Physical review letters}\ }\textbf {\bibinfo {volume} {113}},\ \bibinfo
  {pages} {245702} (\bibinfo {year} {2014})}\BibitemShut {NoStop}%
\bibitem [{\citenamefont {Wu}\ \emph {et~al.}(2015)\citenamefont {Wu},
  \citenamefont {Iwashita},\ and\ \citenamefont {Egami}}]{Wu15}%
  \BibitemOpen
  \bibfield  {author} {\bibinfo {author} {\bibfnamefont {B.}~\bibnamefont
  {Wu}}, \bibinfo {author} {\bibfnamefont {T.}~\bibnamefont {Iwashita}}, \ and\
  \bibinfo {author} {\bibfnamefont {T.}~\bibnamefont {Egami}},\ }\href@noop {}
  {\bibfield  {journal} {\bibinfo  {journal} {Physical Review E}\ }\textbf
  {\bibinfo {volume} {91}},\ \bibinfo {pages} {032301} (\bibinfo {year}
  {2015})}\BibitemShut {NoStop}%
\bibitem [{\citenamefont {Lema{\^\i}tre}(2015)}]{Lemaitre15}%
  \BibitemOpen
  \bibfield  {author} {\bibinfo {author} {\bibfnamefont {A.}~\bibnamefont
  {Lema{\^\i}tre}},\ }\href@noop {} {\bibfield  {journal} {\bibinfo  {journal}
  {The Journal of chemical physics}\ }\textbf {\bibinfo {volume} {143}},\
  \bibinfo {pages} {164515} (\bibinfo {year} {2015})}\BibitemShut {NoStop}%
\bibitem [{\citenamefont {Jensen}\ \emph {et~al.}(2014)\citenamefont {Jensen},
  \citenamefont {Weitz},\ and\ \citenamefont {Spaepen}}]{Jensen14}%
  \BibitemOpen
  \bibfield  {author} {\bibinfo {author} {\bibfnamefont {K.}~\bibnamefont
  {Jensen}}, \bibinfo {author} {\bibfnamefont {D.~A.}\ \bibnamefont {Weitz}}, \
  and\ \bibinfo {author} {\bibfnamefont {F.}~\bibnamefont {Spaepen}},\
  }\href@noop {} {\bibfield  {journal} {\bibinfo  {journal} {Physical Review
  E}\ }\textbf {\bibinfo {volume} {90}},\ \bibinfo {pages} {042305} (\bibinfo
  {year} {2014})}\BibitemShut {NoStop}%
\bibitem [{\citenamefont {Illing}\ \emph {et~al.}(2016)\citenamefont {Illing},
  \citenamefont {Fritschi}, \citenamefont {Hajnal}, \citenamefont {Klix},
  \citenamefont {Keim},\ and\ \citenamefont {Fuchs}}]{Illing16}%
  \BibitemOpen
  \bibfield  {author} {\bibinfo {author} {\bibfnamefont {B.}~\bibnamefont
  {Illing}}, \bibinfo {author} {\bibfnamefont {S.}~\bibnamefont {Fritschi}},
  \bibinfo {author} {\bibfnamefont {D.}~\bibnamefont {Hajnal}}, \bibinfo
  {author} {\bibfnamefont {C.}~\bibnamefont {Klix}}, \bibinfo {author}
  {\bibfnamefont {P.}~\bibnamefont {Keim}}, \ and\ \bibinfo {author}
  {\bibfnamefont {M.}~\bibnamefont {Fuchs}},\ }\href@noop {} {\bibfield
  {journal} {\bibinfo  {journal} {Physical review letters}\ }\textbf {\bibinfo
  {volume} {117}},\ \bibinfo {pages} {208002} (\bibinfo {year}
  {2016})}\BibitemShut {NoStop}%
\bibitem [{\citenamefont {Le~Bouil}\ \emph {et~al.}(2014)\citenamefont
  {Le~Bouil}, \citenamefont {Amon}, \citenamefont {McNamara},\ and\
  \citenamefont {Crassous}}]{Le-Bouil14}%
  \BibitemOpen
  \bibfield  {author} {\bibinfo {author} {\bibfnamefont {A.}~\bibnamefont
  {Le~Bouil}}, \bibinfo {author} {\bibfnamefont {A.}~\bibnamefont {Amon}},
  \bibinfo {author} {\bibfnamefont {S.}~\bibnamefont {McNamara}}, \ and\
  \bibinfo {author} {\bibfnamefont {J.}~\bibnamefont {Crassous}},\ }\href@noop
  {} {\bibfield  {journal} {\bibinfo  {journal} {Physical review letters}\
  }\textbf {\bibinfo {volume} {112}},\ \bibinfo {pages} {246001} (\bibinfo
  {year} {2014})}\BibitemShut {NoStop}%
\bibitem [{\citenamefont {Maier}\ \emph {et~al.}(2017)\citenamefont {Maier},
  \citenamefont {Zippelius},\ and\ \citenamefont {Fuchs}}]{Maier17}%
  \BibitemOpen
  \bibfield  {author} {\bibinfo {author} {\bibfnamefont {M.}~\bibnamefont
  {Maier}}, \bibinfo {author} {\bibfnamefont {A.}~\bibnamefont {Zippelius}}, \
  and\ \bibinfo {author} {\bibfnamefont {M.}~\bibnamefont {Fuchs}},\
  }\href@noop {} {\bibfield  {journal} {\bibinfo  {journal} {Physical review
  letters}\ }\textbf {\bibinfo {volume} {119}},\ \bibinfo {pages} {265701}
  (\bibinfo {year} {2017})}\BibitemShut {NoStop}%
\bibitem [{\citenamefont {Roux}\ and\ \citenamefont {Combe}(2010)}]{Roux10}%
  \BibitemOpen
  \bibfield  {author} {\bibinfo {author} {\bibfnamefont {J.}~\bibnamefont
  {Roux}}\ and\ \bibinfo {author} {\bibfnamefont {G.}~\bibnamefont {Combe}},\
  }in\ \href@noop {} {\emph {\bibinfo {booktitle} {Proceedings of the
  IUTAM-ISIMM Symposium on Mathematical Modeling and Physical Instances of
  Granular Flow}}},\ Vol.\ \bibinfo {volume} {1227}\ (\bibinfo  {publisher}
  {AIP},\ \bibinfo {year} {2010})\ pp.\ \bibinfo {pages} {260--270}\BibitemShut
  {NoStop}%
\bibitem [{\citenamefont {Schreck}\ \emph {et~al.}(2011)\citenamefont
  {Schreck}, \citenamefont {Bertrand}, \citenamefont {O'Hern},\ and\
  \citenamefont {Shattuck}}]{Schreck11}%
  \BibitemOpen
  \bibfield  {author} {\bibinfo {author} {\bibfnamefont {C.~F.}\ \bibnamefont
  {Schreck}}, \bibinfo {author} {\bibfnamefont {T.}~\bibnamefont {Bertrand}},
  \bibinfo {author} {\bibfnamefont {C.~S.}\ \bibnamefont {O'Hern}}, \ and\
  \bibinfo {author} {\bibfnamefont {M.~D.}\ \bibnamefont {Shattuck}},\ }\href
  {\doibase 10.1103/PhysRevLett.107.078301} {\bibfield  {journal} {\bibinfo
  {journal} {Phys. Rev. Lett.}\ }\textbf {\bibinfo {volume} {107}},\ \bibinfo
  {pages} {078301} (\bibinfo {year} {2011})}\BibitemShut {NoStop}%
\bibitem [{\citenamefont {Lema{\^\i}tre}(2017)}]{Lemaitre17}%
  \BibitemOpen
  \bibfield  {author} {\bibinfo {author} {\bibfnamefont {A.}~\bibnamefont
  {Lema{\^\i}tre}},\ }\href@noop {} {\bibfield  {journal} {\bibinfo  {journal}
  {Physical Review E}\ }\textbf {\bibinfo {volume} {96}},\ \bibinfo {pages}
  {052101} (\bibinfo {year} {2017})}\BibitemShut {NoStop}%
\bibitem [{\citenamefont {Liu}\ \emph {et~al.}(2010)\citenamefont {Liu},
  \citenamefont {Nagel}, \citenamefont {van Saarloos},\ and\ \citenamefont
  {Wyart}}]{Liu10}%
  \BibitemOpen
  \bibfield  {author} {\bibinfo {author} {\bibfnamefont {A.~J.}\ \bibnamefont
  {Liu}}, \bibinfo {author} {\bibfnamefont {S.~R.}\ \bibnamefont {Nagel}},
  \bibinfo {author} {\bibfnamefont {W.}~\bibnamefont {van Saarloos}}, \ and\
  \bibinfo {author} {\bibfnamefont {M.}~\bibnamefont {Wyart}},\ }\enquote
  {\bibinfo {title} {The jamming scenario: an introduction and outlook},}\ in\
  \href@noop {} {\emph {\bibinfo {booktitle} {Dynamical heterogeneities in
  glasses, colloids, and granular media}}},\ \bibinfo {editor} {edited by\
  \bibinfo {editor} {\bibnamefont {L.Berthier}}, \bibinfo {editor}
  {\bibfnamefont {G.}~\bibnamefont {Biroli}}, \bibinfo {editor} {\bibfnamefont
  {J.}~\bibnamefont {Bouchaud}}, \bibinfo {editor} {\bibfnamefont
  {L.}~\bibnamefont {Cipeletti}}, \ and\ \bibinfo {editor} {\bibfnamefont
  {W.}~\bibnamefont {van Saarloos}}}\ (\bibinfo  {publisher} {Oxford University
  Press},\ \bibinfo {address} {Oxford},\ \bibinfo {year} {2010})\BibitemShut
  {NoStop}%
\bibitem [{\citenamefont {Wyart}(2012)}]{Wyart12}%
  \BibitemOpen
  \bibfield  {author} {\bibinfo {author} {\bibfnamefont {M.}~\bibnamefont
  {Wyart}},\ }\href {\doibase 10.1103/PhysRevLett.109.125502} {\bibfield
  {journal} {\bibinfo  {journal} {Phys. Rev. Lett.}\ }\textbf {\bibinfo
  {volume} {109}},\ \bibinfo {pages} {125502} (\bibinfo {year}
  {2012})}\BibitemShut {NoStop}%
\bibitem [{\citenamefont {DeGiuli}(2018)}]{atmp_DeGiuli18_long}%
  \BibitemOpen
  \bibfield  {author} {\bibinfo {author} {\bibfnamefont {E.}~\bibnamefont
  {DeGiuli}},\ }\href@noop {} {\enquote {\bibinfo {title} {Edwards field theory
  for glasses and granular matter},}\ } (\bibinfo {year} {2018}),\ \bibinfo
  {note} {to appear in Phys. Rev. E}\BibitemShut {NoStop}%
\bibitem [{\citenamefont {Bi}\ \emph {et~al.}(2015)\citenamefont {Bi},
  \citenamefont {Henkes}, \citenamefont {Daniels},\ and\ \citenamefont
  {Chakraborty}}]{Bi15}%
  \BibitemOpen
  \bibfield  {author} {\bibinfo {author} {\bibfnamefont {D.}~\bibnamefont
  {Bi}}, \bibinfo {author} {\bibfnamefont {S.}~\bibnamefont {Henkes}}, \bibinfo
  {author} {\bibfnamefont {K.~E.}\ \bibnamefont {Daniels}}, \ and\ \bibinfo
  {author} {\bibfnamefont {B.}~\bibnamefont {Chakraborty}},\ }\href@noop {}
  {\bibfield  {journal} {\bibinfo  {journal} {Annu. Rev. Condens. Matter
  Phys.}\ }\textbf {\bibinfo {volume} {6}},\ \bibinfo {pages} {63} (\bibinfo
  {year} {2015})}\BibitemShut {NoStop}%
\bibitem [{\citenamefont {Chakraborty}(2010)}]{Chakraborty10}%
  \BibitemOpen
  \bibfield  {author} {\bibinfo {author} {\bibfnamefont {B.}~\bibnamefont
  {Chakraborty}},\ }\href {\doibase 10.1039/B927435A} {\bibfield  {journal}
  {\bibinfo  {journal} {Soft Matter}\ }\textbf {\bibinfo {volume} {6}},\
  \bibinfo {pages} {2884} (\bibinfo {year} {2010})}\BibitemShut {NoStop}%
\bibitem [{\citenamefont {Lerner}\ and\ \citenamefont
  {Bouchbinder}(2018)}]{Lerner18}%
  \BibitemOpen
  \bibfield  {author} {\bibinfo {author} {\bibfnamefont {E.}~\bibnamefont
  {Lerner}}\ and\ \bibinfo {author} {\bibfnamefont {E.}~\bibnamefont
  {Bouchbinder}},\ }\href@noop {} {\bibfield  {journal} {\bibinfo  {journal}
  {The Journal of chemical physics}\ }\textbf {\bibinfo {volume} {148}},\
  \bibinfo {pages} {214502} (\bibinfo {year} {2018})}\BibitemShut {NoStop}%
\bibitem [{\citenamefont {Goldenberg}\ \emph {et~al.}(2007)\citenamefont
  {Goldenberg}, \citenamefont {Tanguy},\ and\ \citenamefont
  {Barrat}}]{Goldenberg07}%
  \BibitemOpen
  \bibfield  {author} {\bibinfo {author} {\bibfnamefont {C.}~\bibnamefont
  {Goldenberg}}, \bibinfo {author} {\bibfnamefont {A.}~\bibnamefont {Tanguy}},
  \ and\ \bibinfo {author} {\bibfnamefont {J.-L.}\ \bibnamefont {Barrat}},\
  }\href@noop {} {\bibfield  {journal} {\bibinfo  {journal} {EPL (Europhysics
  Letters)}\ }\textbf {\bibinfo {volume} {80}},\ \bibinfo {pages} {16003 
  0295} (\bibinfo {year} {2007})}\BibitemShut {NoStop}%
\bibitem [{\citenamefont {Muskhelishvili}(1963)}]{Muskhelishvili63}%
  \BibitemOpen
  \bibfield  {author} {\bibinfo {author} {\bibfnamefont {N.}~\bibnamefont
  {Muskhelishvili}},\ }\href@noop {} {\emph {\bibinfo {title} {Some basic
  problems of the mathematical theory of elasticity}}}\ (\bibinfo  {publisher}
  {P. Noordhoff},\ \bibinfo {address} {Groningen},\ \bibinfo {year}
  {1963})\BibitemShut {NoStop}%
\bibitem [{\citenamefont {Ball}\ and\ \citenamefont
  {Blumenfeld}(2002)}]{Ball02}%
  \BibitemOpen
  \bibfield  {author} {\bibinfo {author} {\bibfnamefont {R.~C.}\ \bibnamefont
  {Ball}}\ and\ \bibinfo {author} {\bibfnamefont {R.}~\bibnamefont
  {Blumenfeld}},\ }\href@noop {} {\bibfield  {journal} {\bibinfo  {journal}
  {Phys. Rev. Lett.}\ }\textbf {\bibinfo {volume} {88}},\ \bibinfo {pages}
  {115505} (\bibinfo {year} {2002})}\BibitemShut {NoStop}%
\bibitem [{\citenamefont {Henkes}(2009)}]{Henkes09}%
  \BibitemOpen
  \bibfield  {author} {\bibinfo {author} {\bibfnamefont {S.}~\bibnamefont
  {Henkes}},\ }\emph {\bibinfo {title} {A statistical mechanics framework for
  static granular matter}},\ \href@noop {} {Ph.D. thesis},\ \bibinfo  {school}
  {Brandeis U.} (\bibinfo {year} {2009})\BibitemShut {NoStop}%
\bibitem [{\citenamefont {DeGiuli}\ and\ \citenamefont
  {McElwaine}(2011)}]{DeGiuli11}%
  \BibitemOpen
  \bibfield  {author} {\bibinfo {author} {\bibfnamefont {E.}~\bibnamefont
  {DeGiuli}}\ and\ \bibinfo {author} {\bibfnamefont {J.}~\bibnamefont
  {McElwaine}},\ }\href {\doibase 10.1103/PhysRevE.84.041310} {\bibfield
  {journal} {\bibinfo  {journal} {Phys. Rev. E}\ }\textbf {\bibinfo {volume}
  {84}},\ \bibinfo {pages} {041310} (\bibinfo {year} {2011})}\BibitemShut
  {NoStop}%
\bibitem [{\citenamefont {DeGiuli}\ and\ \citenamefont
  {Schoof}(2014)}]{DeGiuli14a}%
  \BibitemOpen
  \bibfield  {author} {\bibinfo {author} {\bibfnamefont {E.}~\bibnamefont
  {DeGiuli}}\ and\ \bibinfo {author} {\bibfnamefont {C.}~\bibnamefont
  {Schoof}},\ }\href {http://stacks.iop.org/0295-5075/105/i=2/a=28001}
  {\bibfield  {journal} {\bibinfo  {journal} {EPL (Europhysics Letters)}\
  }\textbf {\bibinfo {volume} {105}},\ \bibinfo {pages} {28001} (\bibinfo
  {year} {2014})}\BibitemShut {NoStop}%
\bibitem [{\citenamefont {DiDonna}\ and\ \citenamefont
  {Lubensky}(2005)}]{DiDonna05}%
  \BibitemOpen
  \bibfield  {author} {\bibinfo {author} {\bibfnamefont {B.}~\bibnamefont
  {DiDonna}}\ and\ \bibinfo {author} {\bibfnamefont {T.}~\bibnamefont
  {Lubensky}},\ }\href@noop {} {\bibfield  {journal} {\bibinfo  {journal}
  {Physical Review E}\ }\textbf {\bibinfo {volume} {72}},\ \bibinfo {pages}
  {066619} (\bibinfo {year} {2005})}\BibitemShut {NoStop}%
\bibitem [{\citenamefont {Gurtin}(1963)}]{Gurtin63}%
  \BibitemOpen
  \bibfield  {author} {\bibinfo {author} {\bibfnamefont {M.~E.}\ \bibnamefont
  {Gurtin}},\ }\href@noop {} {\bibfield  {journal} {\bibinfo  {journal}
  {Archive for Rational Mechanics and Analysis}\ }\textbf {\bibinfo {volume}
  {13}},\ \bibinfo {pages} {321} (\bibinfo {year} {1963})}\BibitemShut
  {NoStop}%
\bibitem [{\citenamefont {Wang}\ and\ \citenamefont
  {Rutqvist}(2013)}]{Wang13a}%
  \BibitemOpen
  \bibfield  {author} {\bibinfo {author} {\bibfnamefont {Y.}~\bibnamefont
  {Wang}}\ and\ \bibinfo {author} {\bibfnamefont {J.}~\bibnamefont
  {Rutqvist}},\ }\href@noop {} {\bibfield  {journal} {\bibinfo  {journal}
  {Journal of Elasticity}\ }\textbf {\bibinfo {volume} {113}},\ \bibinfo
  {pages} {283} (\bibinfo {year} {2013})}\BibitemShut {NoStop}%
\bibitem [{\citenamefont {Berthier}\ and\ \citenamefont
  {Biroli}(2011)}]{Berthier11b}%
  \BibitemOpen
  \bibfield  {author} {\bibinfo {author} {\bibfnamefont {L.}~\bibnamefont
  {Berthier}}\ and\ \bibinfo {author} {\bibfnamefont {G.}~\bibnamefont
  {Biroli}},\ }\href@noop {} {\bibfield  {journal} {\bibinfo  {journal}
  {Reviews of Modern Physics}\ }\textbf {\bibinfo {volume} {83}},\ \bibinfo
  {pages} {587} (\bibinfo {year} {2011})}\BibitemShut {NoStop}%
\bibitem [{\citenamefont {Tighe}\ \emph {et~al.}(2008)\citenamefont {Tighe},
  \citenamefont {van Eerd},\ and\ \citenamefont {Vlugt}}]{Tighe08}%
  \BibitemOpen
  \bibfield  {author} {\bibinfo {author} {\bibfnamefont {B.~P.}\ \bibnamefont
  {Tighe}}, \bibinfo {author} {\bibfnamefont {A.~R.~T.}\ \bibnamefont {van
  Eerd}}, \ and\ \bibinfo {author} {\bibfnamefont {T.~J.~H.}\ \bibnamefont
  {Vlugt}},\ }\href {\doibase 10.1103/PhysRevLett.100.238001} {\bibfield
  {journal} {\bibinfo  {journal} {Phys. Rev. Lett.}\ }\textbf {\bibinfo
  {volume} {100}},\ \bibinfo {pages} {238001} (\bibinfo {year}
  {2008})}\BibitemShut {NoStop}%
\bibitem [{\citenamefont {Tighe}\ \emph {et~al.}(2010)\citenamefont {Tighe},
  \citenamefont {Snoeijer}, \citenamefont {Vlugt},\ and\ \citenamefont {van
  Hecke}}]{Tighe10a}%
  \BibitemOpen
  \bibfield  {author} {\bibinfo {author} {\bibfnamefont {B.~P.}\ \bibnamefont
  {Tighe}}, \bibinfo {author} {\bibfnamefont {J.~H.}\ \bibnamefont {Snoeijer}},
  \bibinfo {author} {\bibfnamefont {T.~J.~H.}\ \bibnamefont {Vlugt}}, \ and\
  \bibinfo {author} {\bibfnamefont {M.}~\bibnamefont {van Hecke}},\ }\href
  {\doibase 10.1039/B926592A} {\bibfield  {journal} {\bibinfo  {journal} {Soft
  Matter}\ }\textbf {\bibinfo {volume} {6}},\ \bibinfo {pages} {2908} (\bibinfo
  {year} {2010})}\BibitemShut {NoStop}%
\bibitem [{\citenamefont {Tighe}\ and\ \citenamefont {Vlugt}(2010)}]{Tighe10b}%
  \BibitemOpen
  \bibfield  {author} {\bibinfo {author} {\bibfnamefont {B.~P.}\ \bibnamefont
  {Tighe}}\ and\ \bibinfo {author} {\bibfnamefont {T.~J.~H.}\ \bibnamefont
  {Vlugt}},\ }\href {http://stacks.iop.org/1742-5468/2010/i=01/a=P01015}
  {\bibfield  {journal} {\bibinfo  {journal} {J. Stat. Mech.}\ }\textbf
  {\bibinfo {volume} {2010}},\ \bibinfo {pages} {P01015} (\bibinfo {year}
  {2010})}\BibitemShut {NoStop}%
\bibitem [{\citenamefont {Wu}\ and\ \citenamefont {Teitel}(2015)}]{Wu15a}%
  \BibitemOpen
  \bibfield  {author} {\bibinfo {author} {\bibfnamefont {Y.}~\bibnamefont
  {Wu}}\ and\ \bibinfo {author} {\bibfnamefont {S.}~\bibnamefont {Teitel}},\
  }\href@noop {} {\bibfield  {journal} {\bibinfo  {journal} {Physical Review
  E}\ }\textbf {\bibinfo {volume} {92}},\ \bibinfo {pages} {022207} (\bibinfo
  {year} {2015})}\BibitemShut {NoStop}%
\bibitem [{\citenamefont {Blumenfeld}\ and\ \citenamefont
  {Edwards}(2009)}]{Blumenfeld09}%
  \BibitemOpen
  \bibfield  {author} {\bibinfo {author} {\bibfnamefont {R.}~\bibnamefont
  {Blumenfeld}}\ and\ \bibinfo {author} {\bibfnamefont {S.~F.}\ \bibnamefont
  {Edwards}},\ }\href@noop {} {\bibfield  {journal} {\bibinfo  {journal} {The
  Journal of Physical Chemistry B}\ }\textbf {\bibinfo {volume} {113}},\
  \bibinfo {pages} {3981} (\bibinfo {year} {2009})}\BibitemShut {NoStop}%
\bibitem [{\citenamefont {Bililign}\ \emph {et~al.}(2018)\citenamefont
  {Bililign}, \citenamefont {Kollmer},\ and\ \citenamefont
  {Daniels}}]{Bililign18}%
  \BibitemOpen
  \bibfield  {author} {\bibinfo {author} {\bibfnamefont {E.~S.}\ \bibnamefont
  {Bililign}}, \bibinfo {author} {\bibfnamefont {J.~E.}\ \bibnamefont
  {Kollmer}}, \ and\ \bibinfo {author} {\bibfnamefont {K.~E.}\ \bibnamefont
  {Daniels}},\ }\href@noop {} {\enquote {\bibinfo {title} {Protocol-dependence
  and state variables in the force-moment ensemble},}\ } (\bibinfo {year}
  {2018}),\ \bibinfo {note} {arXiv:1802.09641}\BibitemShut {NoStop}%
\bibitem [{\citenamefont {Bertin}\ \emph {et~al.}(2006)\citenamefont {Bertin},
  \citenamefont {Dauchot},\ and\ \citenamefont {Droz}}]{Bertin06}%
  \BibitemOpen
  \bibfield  {author} {\bibinfo {author} {\bibfnamefont {E.}~\bibnamefont
  {Bertin}}, \bibinfo {author} {\bibfnamefont {O.}~\bibnamefont {Dauchot}}, \
  and\ \bibinfo {author} {\bibfnamefont {M.}~\bibnamefont {Droz}},\ }\href@noop
  {} {\bibfield  {journal} {\bibinfo  {journal} {Physical review letters}\
  }\textbf {\bibinfo {volume} {96}},\ \bibinfo {pages} {120601} (\bibinfo
  {year} {2006})}\BibitemShut {NoStop}%
\bibitem [{\citenamefont {Henkes}\ \emph {et~al.}(2007)\citenamefont {Henkes},
  \citenamefont {O'Hern},\ and\ \citenamefont {Chakraborty}}]{Henkes07}%
  \BibitemOpen
  \bibfield  {author} {\bibinfo {author} {\bibfnamefont {S.}~\bibnamefont
  {Henkes}}, \bibinfo {author} {\bibfnamefont {C.~S.}\ \bibnamefont {O'Hern}},
  \ and\ \bibinfo {author} {\bibfnamefont {B.}~\bibnamefont {Chakraborty}},\
  }\href {\doibase 10.1103/PhysRevLett.99.038002} {\bibfield  {journal}
  {\bibinfo  {journal} {Phys. Rev. Lett.}\ }\textbf {\bibinfo {volume} {99}},\
  \bibinfo {pages} {038002} (\bibinfo {year} {2007})}\BibitemShut {NoStop}%
\bibitem [{\citenamefont {Puckett}\ and\ \citenamefont
  {Daniels}(2013)}]{Puckett13}%
  \BibitemOpen
  \bibfield  {author} {\bibinfo {author} {\bibfnamefont {J.~G.}\ \bibnamefont
  {Puckett}}\ and\ \bibinfo {author} {\bibfnamefont {K.~E.}\ \bibnamefont
  {Daniels}},\ }\href@noop {} {\bibfield  {journal} {\bibinfo  {journal}
  {Physical Review Letters}\ }\textbf {\bibinfo {volume} {110}},\ \bibinfo
  {pages} {058001} (\bibinfo {year} {2013})}\BibitemShut {NoStop}%
\bibitem [{\citenamefont {Barrat}\ \emph {et~al.}(2000)\citenamefont {Barrat},
  \citenamefont {Kurchan}, \citenamefont {Loreto},\ and\ \citenamefont
  {Sellitto}}]{Barrat00}%
  \BibitemOpen
  \bibfield  {author} {\bibinfo {author} {\bibfnamefont {A.}~\bibnamefont
  {Barrat}}, \bibinfo {author} {\bibfnamefont {J.}~\bibnamefont {Kurchan}},
  \bibinfo {author} {\bibfnamefont {V.}~\bibnamefont {Loreto}}, \ and\ \bibinfo
  {author} {\bibfnamefont {M.}~\bibnamefont {Sellitto}},\ }\href@noop {}
  {\bibfield  {journal} {\bibinfo  {journal} {Physical review letters}\
  }\textbf {\bibinfo {volume} {85}},\ \bibinfo {pages} {5034} (\bibinfo {year}
  {2000})}\BibitemShut {NoStop}%
\bibitem [{\citenamefont {Biroli}\ and\ \citenamefont
  {Kurchan}(2001)}]{Biroli01a}%
  \BibitemOpen
  \bibfield  {author} {\bibinfo {author} {\bibfnamefont {G.}~\bibnamefont
  {Biroli}}\ and\ \bibinfo {author} {\bibfnamefont {J.}~\bibnamefont
  {Kurchan}},\ }\href@noop {} {\bibfield  {journal} {\bibinfo  {journal}
  {Physical Review E}\ }\textbf {\bibinfo {volume} {64}},\ \bibinfo {pages}
  {016101} (\bibinfo {year} {2001})}\BibitemShut {NoStop}%
\bibitem [{\citenamefont {Zee}(2010)}]{Zee10}%
  \BibitemOpen
  \bibfield  {author} {\bibinfo {author} {\bibfnamefont {A.}~\bibnamefont
  {Zee}},\ }\href@noop {} {\emph {\bibinfo {title} {Quantum field theory in a
  nutshell}}}\ (\bibinfo  {publisher} {Princeton university press},\ \bibinfo
  {year} {2010})\BibitemShut {NoStop}%
\bibitem [{\citenamefont {Zinn-Justin}(1996)}]{Zinn-Justin96}%
  \BibitemOpen
  \bibfield  {author} {\bibinfo {author} {\bibfnamefont {J.~.~X.}\ \bibnamefont
  {Zinn-Justin}},\ }\href@noop {} {\emph {\bibinfo {title} {Quantum field
  theory and critical phenomena}}}\ (\bibinfo  {publisher} {Clarendon Press},\
  \bibinfo {year} {1996})\BibitemShut {NoStop}%
\bibitem [{\citenamefont {Sadd}(2009)}]{Sadd09}%
  \BibitemOpen
  \bibfield  {author} {\bibinfo {author} {\bibfnamefont {M.~H.}\ \bibnamefont
  {Sadd}},\ }\href@noop {} {\emph {\bibinfo {title} {Elasticity - Theory,
  Applications, and Numerics (2nd Edition)}}}\ (\bibinfo  {publisher}
  {Elsevier},\ \bibinfo {year} {2009})\BibitemShut {NoStop}%
\end{thebibliography}%

\end{document}